\documentclass[a4paper]{article}
\usepackage{interspeech_04,amssymb,amsmath,epsfig,times,multirow}
\ninept

\newcommand{\enumbegin}
  { \begin{list}{$\bullet$}
      { \setlength{\itemsep}{0pt}
	\setlength{\parsep}{0pt}
	\setlength{\topsep}{0pt}
	\setlength{\partopsep}{0pt}
	\setlength{\leftmargin}{1.5em}
	\setlength{\labelwidth}{1em}
	\setlength{\labelsep}{0.5em} } }
\newcommand{\enumend}
  { \end{list} }

\title{Detecting User Engagement in Everyday Conversations}

\twoauthors 
{Chen Yu}
{Department of Computer Science\\
University of Rochester\\
Rochester, NY  14627  USA\\
{\small \tt yu@cs.rochester.edu}}
{Paul M. Aoki and Allison Woodruff}
{Palo Alto Research Center\\
3333 Coyote Hill Road\\
Palo Alto, CA  94304  USA\\
{\small \tt \{aoki,woodruff\}@acm.org}}

\begin{document}
\maketitle
\begin{abstract}
  This paper presents a novel application of speech emotion
  recognition: estimation of the level of conversational engagement
  between users of a voice communication system. We begin by
  using machine learning techniques, such as the support vector
  machine (SVM), to classify users' emotions as expressed in
  individual utterances. However, this alone fails to model the
  temporal and interactive aspects of conversational engagement.  We
  therefore propose the use of a multilevel structure based on coupled
  hidden Markov models (HMM) to estimate engagement levels in
  continuous natural speech. The first level is comprised of SVM-based
  classifiers that recognize emotional states, which could be (e.g.)
  discrete emotion types or arousal/valence levels. A high-level HMM
  then uses
  these emotional states as input, estimating users' engagement in
  conversation by decoding the internal states of the HMM. We report
  experimental results obtained by applying our algorithms to the LDC
  {\sc Emotional Prosody} and {\sc CallFriend} speech corpora.
\end{abstract}

\section{Introduction}
Wireless personal communication systems, such as the mobile phone
network, continue to advance rapidly in terms of technology as well as
breadth of deployment.  However, even with all of these advances,
almost all mobile voice communication still takes place using the
telephone call model that originated in the 19th Century.  Field
studies of current wireless communication technologies in use
(e.g.,~\cite{woodruff03}) show that users gravitate toward
communication patterns that are more lightweight, dynamic and
spontaneous than those seen with non-mobile technologies.  We are
designing new voice communication systems that are intended to fit
better into these new patterns of use.  In each case, the
communication system uses machine learning techniques to model the
state of social interaction in the voice channel, and then adapts the
channel to facilitate interaction.

In this paper, we describe the machine learning component of one such
adaptive communication system.  The system relies on estimates of the
level of the users' {\em engagement} in an ongoing remote
conversation.  One use of these engagement estimates is to increase or
decrease the ``richness'' of a communication session in an automatic
and seamless way.  For example, if two users are speaking in a
push-to-talk (half-duplex audio) session and become highly engaged,
the system could automatically switch over to a telephony (duplex
audio) connection.  Similarly, if the two participants become even
more engaged in the telephone conversation, the system could then add
a video channel.  (A longer discussion can be found
in~\cite{woodruff03}.)

One can attempt to capture an instantaneous notion of conversational
participants' feelings by analyzing speech as it passes through the
voice channel.  In addition to carrying linguistic information, the
manner in which speech is delivered provides many acoustic cues
indicating the speaker's emotions and attitudes toward the topic, the
dialogue partner, the situation, etc.  In this work, we attempt to develop
a computer system to extract such non-linguistic information from
users' speech.  In doing so, we build on previous work relating
emotions to speech.  The correlations between acoustic features (e.g.,
prosody) and emotional states have been studied in speech production
and phonetics for many years (for a review, see~\cite{scherer03}).
Recently, there has been growing interest in automatic emotion
recognition from speech, with diverse approaches being applied.
For example,
Dellaert {\em et al.}~\cite{dellaert96} implemented a method based on
the majority voting of subspace specialists to classify acted spoken
utterances into four emotion types;
Batliner {\em et al.}~\cite{batliner00} provided a comparative study
of recognizing two emotion types, ``neutral'' and ``anger,'' expressed by
actors and naive subjects;
Schr\"{o}der {\em et al.}~\cite{schroeder01} analyzed the correlation
between various acoustic features and three abstract emotional
characteristics in a collection of broadcast media recordings;
and Lee {\em et al.}~\cite{lee02} combined acoustic features with the
recognition of emotionally salient words to categorize utterances as
negative or non-negative.
A good review of automatic emotion recognition can be found
in~\cite{cowie01}.

\begin{figure*}[tbp]
\centerline{\epsfig{figure=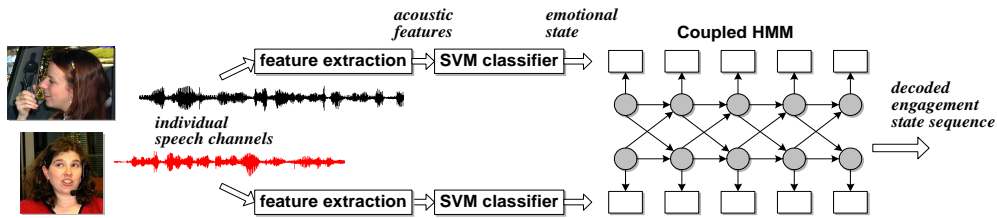,height=30mm,width=138mm}}
\caption{\em Overview of a multilevel structure for engagement detection.}
\label{fig:overview}
\end{figure*}

The work reported here is novel in two ways.  First, we adapt prior work on
emotion recognition to the estimation of conversational engagement.
Second, we formulate the problem of estimating engagement from speech
as one that has multiple inputs -- that is, it considers acoustic,
temporal, and interactional information.  This formulation follows
from a recognition that conversation is a sequentially-organized,
social process, and that a speaker's display of affect (including the
acoustic cues in speech) is, in large part, part of a trajectory of
conversational actions that includes all of the participants.
Empirical research in conversation analysis has (qualitatively)
demonstrated many linkages between emotional displays and
conversational engagement (see, e.g.,~\cite{selting94,goodwin00}).

The central idea, then, is that an operational model of engagement in
everyday conversation should directly reflect the fact that a
participant's current engagement state is influenced by his/her
previous engagement state (temporal continuity), his/her current
emotional state (as reflected by affective display, which is speech
here), and the other participants' engagement states (as expressed
through interaction).  Aside from theoretical aspects, there are two
(potential) practical advantages of considering these multiple inputs.
First, we may be able to get better accuracy because we can compensate
for transient noise in one input in the integration process. Second,
in the case that some information is not available, we may still be
able to compute users' engagement based on partial information. For
example, when a given user listens silently to the current speaker's
talk, a method depending solely on the acoustic features of speech
cannot estimate the listener's engagement.  In our method, however, we
can use the listener's previous engagement level and the speaker's
current engagement level to estimate the listener's current engagement
level.

In light of the above analysis, we propose a multilevel structure
using support vector machine (SVM) and hidden Markov model (HMM)
techniques as shown in Figure~\ref{fig:overview}.  The first level of
the architecture is comprised of SVM classifiers that use acoustic
features as input and predict emotional states of users, which could
be (e.g.)  discrete emotion types or arousal/valence levels. Those
emotional states are
utilized as input to the higher level HMM.  This models a
user's emotional state and engagement in conversation as a dynamic,
continous process.  In addition, we apply the
coupled HMM (CHMM) technique to capture joint behavior of
participants and model the influence of individual participants on
each other. In this way, the method decodes users' engagement states
in conversations by seamlessly integrating low-level prosodic,
temporal and cross-participant cues.  In the rest of the paper,
Section~\ref{sec:emotion} presents our method of speech emotion recognition and
Section~\ref{sec:engagement} describes engagement detection based on CHMM. The
experimental results are reported and discussed in Section~\ref{sec:results}.

\section{Speech emotion recognition}
\label{sec:emotion}
The first level classifies emotions through certain attributes of
spoken communication.  As in any classification problem, we need to
define appropriate target classes and then find features that
can be used to classify accurately.  
We discuss each in turn.

\noindent {\bf Classes.}
In the literature on spoken emotion recognition, there are two main
approaches to classifying emotional states~\cite{cowie01,scherer03}.
In this work, we apply both.

A given emotional state can be characterized using a set of
common-sense labels, or {\em discrete emotion types}, such as
``happy'' or ``sad.''  The various spoken emotion recognition studies
often use very different sets, both for theoretical and practical
reasons (e.g., only a few discrete emotion types may be of interest in
a given application).  In this work, we focus on seven discrete
emotion types: hot anger, panic, sadness, happy, interest, boredom and
no emotion.

Emotions can also be characterized in terms of {\em continuous
dimensions}.  The dimensions define a space in which any given
emotional state can be located; if the dimensions are discretized into
{\em levels}, producing an multidimensional array of possible
characterizations, we define a finite set of classes.  In this work,
we explore the use of the two most commonly considered dimensions,
{\em arousal} and {\em valence}.  Arousal refers to the degree of
intensity of affect and ranges from sleep to excitement.  Valence
describes the pleasantness of the stimuli, such as positive (happy)
and negative (sad).

We now turn to features.
The practical motivations of this work require us to build a
speaker-independent emotion detection system
that has to deal with 
speaker variability in speech. People's willingness
to display emotional responses and the way that they convey affective
messages using speech vary widely across individuals.
Thus, we need to
find a set of robust features that are not only closely correlated
with emotional categories but also invariant across different
speakers.

\noindent {\bf Candidate features.}
We first segment continuous speech into spoken utterances. Then we use the
{\sc Praat} software package to extract the prosodic and energy profiles of
each spoken utterance, which carry a large amount of information of
emotion. Next, seven kinds of acoustic features are extracted from
each spoken utterance:

\enumbegin
\item Fundamental frequency (F0/pitch): mean, maximum, minimum,
  standard deviation, range, 25-percentile, and 75-percentile.
\item Derivative of pitch (rate of change): mean, maximum, minimum, standard
  deviation, and range of derivative; mean and standard
  deviation of absolute derivative. 
\item Duration of pitch: ratio of duration of voiced and unvoiced
regions, mean of frames of voiced regions, standard deviation of frames
of voiced regions, number of voiced regions, ratio of frames of voiced
and unvoiced regions, maximal frames of voiced duration and mean of the
maximum pitch in every region. 
\item Energy: mean, standard deviation, maximum, median, and energy in
  frequency bands ($<$200Hz, 200-300Hz, 300-500Hz, 500-1KHz, 1k-2kHz and
  $>$2kHz).  
\item Derivative of energy (rate of change):  mean, standard deviation, maximum,
  median, and minimum.   
\item Duration of energy in non-silent regions: mean of the number of
  frames, standard deviation of the number of frames,  ratio of
  non-silent frames, and maximum of the number of frames. 
\item Formants: first three formant frequencies (F1, F2, F3), and their
  bandwidths.
\enumend

\noindent {\bf Feature selection.}
Now we have multidimensional features for each spoken utterance. The
curse of dimensionality in high-dimensional classification is
well-known in machine learning, which indicates that pruning the
irrelevant features holds more promise for a generalized
classification. We transformed the original feature space into a lower
dimensional space using the {\sc ReliefF} algorithm for feature
selection.

As mentioned above, we want to develop a speaker-independent emotion
recognition system which needs to deal with speaker variation.
However, since the acoustic features of male and female speakers
differ considerably, we divided users by gender and used different
feature sets for the two groups. For example, the top seven features for
arousal level classification are as follows:

\enumbegin
\item Male: range of F2, range of pitch, maximum of pitch,
  energy $>$2000Hz, maximum of voiced durations, standard deviation of
  derivative of energy, and maximum of energy.
\item Female: mean of pitch, range of derivative of pitch, mean
  duration of voiced regions, energy $<$200Hz, ratio of number of
  silent vs. non-silent frames, maximum of pitch, and maximum of energy.
\enumend

\noindent {\bf Classification.}
We used the SVM algorithm to classify acoustic features, which
projects feature vectors in a higher dimension and constructs a
hyperplane as the decision surface so that the margin of separation
between positive and negative examples is maximized. Specifically, we
used $c-$SVM with polynomial kernels and combined several pairwise
binary SVM classifiers to build multi-class classification by using
the one-against-all method~\cite{lin02}.

\section{User engagement recognition}
\label{sec:engagement}
User engagement measures the commitment to interaction. Thus, it
describes how much a participant is interested in and attentive to a
conversation. To detect this internal state of a participant, the
central idea of our method is that user engagement in conversations
has temporal characteristics that cannot be inferred solely from
low-level perceptual features,
such as acoustic features. However, some user states, such as arousal
levels, not only can be directly estimated from acoustic features but
also are inherently correlated with user engagement. In light of this,
we propose a multilevel structure in which low-level classifiers use
acoustic features extracted from raw speech signals to categorize
spoken utterances in several dimensions (e.g., arousal levels). The
outputs of those classifiers are then used as the observations of the
high-level HMM. The HMM is comprised of five hidden states which
correspond to the degree of engagement in conversations and model
temporal continuity of user engagement.

In addition, we applied a CHMM to describe the mutual influence of
participants' engagement states.  In the CHMM, each chain has five
hidden states corresponding to engagement levels. The observations are
emotional states as received from the low-level classifier. For
example, in the specific CHMM used in Sec.~\ref{sec:engstudy}, the
observations are arousal levels.

\noindent {\bf Training.}
Given the sequences of arousal levels and engagement levels of
participants, the CHMM training procedure needs to estimate three
kinds of probabilities:

\enumbegin
\item $p(o_{i}|s_{j})$ is the probability of observing arousal level
  $i$ in state $j$ which is a multinomial distribution and can be
  learned by simply counting the expected frequency of arousal levels
  in state $j$.
\item $p(s_{j}^{m}|s_{i}^{m})$ is the transition probability of taking the
  transition from state $i$ to state $j$ in chain $m$.
\item $p(s_{j}^{m}|s_{i}^{n})$ is the cross-participant influence
  probability of taking the transition to state $j$ in chain $m$ being
  in state $i$ in chain $n$.
\enumend

\noindent Note that currently the observations are discrete values
(discretized arousal levels, etc.) and are modeled by multinomial
distributions. (We could use Gaussian mixture models for continuous
CHMM observations if we were to use low-level classifiers that
provided probabilistic categories.)

\noindent {\bf Testing.}
In the testing phase, acoustic features are fed into the low-level SVM
classifiers and the output arousal states are fed into the high-level
CHMM. The decoded state sequences of the CHMM are obtained using the Viterbi
algorithm and indicate the engagement states of participants.
Formally, assume that the CHMM consists of two chains corresponding to two
participants in a conversation, and let $s^{1}_{t}$ and $s^{2}_{t}$ be
the engagement states of participant 1 and participant 2 at time $t$
separately.  $o_{t}^{1}$ and $o_{t}^{2}$ are the observations (arousal
levels, etc.) of participants. The model predicts the current state
$s^{1}_{t}$ based on its own previous state $s^{1}_{t-1}$,
cross-channel influence $s^{2}_{t-1}$ and new observation of arousal
level $o_{t}^{1}$.  Specifically, the probability of the combination
of two participant's states are as follows:

\begin{table}[ht]
\vspace{1mm}
\centering
\begin{tabular}{r @{~~}c @{~~}l}
{$p(s^{1}_{t}, s^{2}_{t})$}
 & {$=$}
 & {$p(s^{1}_{t}|s^{1}_{t-1})p(s^{2}_{t}|s^{2}_{t-1})p(s^{1}_{t}|s^{2}_{t-1})p(s^{2}_{t}|s^{1}_{t-1})$}
 \\
{}
 & {}
 & {$p(o^{1}_{t}|s_{t}^{1})p(o^{2}_{t}|s_{t}^{2})$}
\end{tabular}
\end{table}

\noindent In this way, our method is able to estimate two participants'
engagement states simultaneously given raw speech data in the
conversation.

\section{Experiments and results}
\label{sec:results}
The evaluation of computational emotion recognition is
challenging for several reasons. First, data from real-life scenarios
is difficult to acquire; much of the research on emotion in speech is
based on recordings of actors/actresses who simulate specific
emotional states.  Second, emotional categories are quite ambiguous in
their definitions, and different researchers propose different sets of
categories. Third, even when there is agreement on a clear definition
of emotion, labeling emotional speech is not straightforward. In a
conversation, a speaker can be thought of as encoding his/her emotions
in speech and listeners can be thought of as decoding the emotional
information from speech.  However, the speaker and listeners may not
agree on the emotion expressed or perceived in an
utterance. Similarly, different listeners may infer different
emotional states given the same utterance. All of these factors make
data collection, evaluation of emotion, and engagement recognition
much more complicated compared to other statistical pattern
recognition problems, such as visual object recognition and text
mining, in which ground truth can be clearly determined.

\subsection{Data and data coding}
In this work, we used two English-language speech corpora
obtained from the Linguistic Data Consortium (LDC):

The LDC {\sc Emotional Prosody} corpus was produced by 
professional actors/actresses expressing 14 discrete emotion
types. There are approximately 25 spoken utterances per discrete
emotion type.  In our experiments, we focused on seven discrete
emotion types that are most important in our application: hot anger,
panic, sadness, happy, interest, boredom and no emotion. In the
experiments reported here, half of the utterances were used for
training and the other half for testing.

The LDC {\sc CallFriend} corpus was collected by the consensual
recording of social telephone conversations between friends. We
selected four dialogues that contained a range of affect and extracted
usable subsets of approximately 10 minutes from each.  Segmenting the
subsets into utterances produced a total of 1011 utterances from four
female speakers and 877 utterances from four male speakers.  Five
labelers were asked to listen to the individual utterances and provide
four separate labels for each utterance: discrete emotion type as a
categorical value, and numerical values (on a discretized 1--5 scale)
for each of arousal, valence and engagement.  We based the final
labels for each utterance on the consensus of all the labelers.
Again, these experiments used half of the utterances for training
and half for testing.

\subsection{Results of emotion recognition in utterances}
Table~\ref{tab:svm} provides examples of the SVM recognition accuracy
for emotional states of individual spoken utterances.  The table
illustrates how accuracy varied with the classifier type (i.e.,
varying numbers of discrete emotion types and arousal/valence levels),
whether the classifier was trained/tested using data in a
speaker-dependent (SD) or speaker-independent (SI) manner, and whether
the classifer was trained/tested on data from the {\sc Emotional
Prosody} (EP) or {\sc CallFriend} (CF) corpora.  For example, with
regard to discrete emotion types, recognition rates for 5 types
(comparison $(a)$) show that, for the same feature extraction and
machine learning method, the performance with acted speech (EP) in SD
mode is better than that with spontaneous speech (CF) in SI mode
($75\%$ {\em vs.} $51\%$).  Given that many studies of speech emotion
recognition do focus on acted speech and speaker-dependent
recognition, this comparison indicates that results under such
artificial conditions cannot be assumed to generalize to a real-life,
speaker-independent scenario. Looking at both types of classifiers,
the difference between 5 and 7 discrete emotion types (see $(b)$) and
between 3 and 5 arousal levels (see $(c)$) illustrate that 
increasing the number of classes can have a negative effect on the
classification accuracy. Finally, we see two interesting results
regarding emotional dimensions alone.  First, recognition rates of
valence levels are not as good as those obtained for arousal (see
$(d)$), which is consistent in principle with the related
psychological studies~\cite{schroeder01}.  Second, the accuracy at
recognizing arousal levels (see $(c)$) is reasonably good using
spontaneous speech data and SI mode.

\begin{table}[ht]
\centering
\caption{\label{tab:svm}{\em SVM classification accuracy on utterances.}}
\begin{tabular}{|l|l|c|c|@{}l @{}l}
\cline{3-4}
  \multicolumn{1}{l}{\bf classifier} &
  {\bf mode} & 
  {\bf EP} & 
  {\bf CF} &
  {} &
  {} \\
\cline{1-4}
  7 discrete types &
  SD &
  {$69\%$} & 
  {-} &
  \multirow{2}{*}{$\left. \begin{array}{@{}l}{}\\{}\end{array}\right\}(b)$} &
  {} \\
\cline{1-4}
  5 discrete types &
  SD &
  {$75\%$} & 
  {$62\%$} &
  {} &
  \multirow{2}{*}{$\left. \begin{array}{@{}l}{}\\{}\end{array}\right\}(a)$} \\
\cline{1-4}
  5 discrete types &
  SI &
  {$60\%$} & 
  {$51\%$} &
  {} &
  {} \\
\cline{1-4}
  5 arousal levels &
  SI &
  {-} & 
  {$58\%$} &
  \multirow{2}{*}{$\left. \begin{array}{@{}l}{}\\{}\end{array}\right\}(c)$} &
  {} \\
\cline{1-4}
  3 arousal levels &
  SI &
  {-} &
  {$67\%$} &
  {} &
  \multirow{2}{*}{$\left. \begin{array}{@{}l}{}\\{}\end{array}\right\}(d)$} \\
\cline{1-4}
  3 valence levels &
  SI &
  {-} &
  {$54\%$} &
  {} &
  {} \\
\cline{1-4}
\end{tabular}
\end{table}

\subsection{Results of engagement detection in continuous speech}
\label{sec:engstudy}
Table~\ref{tab:table2} shows the results of applying three
speaker-independent methods to the assessment of users' engagement
states on a 1--5 scale (in which a random choice of state would
produce $20\%$ accuracy). We first trained an SVM classifier to
categorize spoken utterances directly, based only on prosodic features, and
obtained $47\%$ accuracy.
A much better result ($61\%$) was achieved by using the multilevel
structure. The
improvement indicates that low-level prosodic features in speech are
only incomplete indicators of users' engagement states and that we
benefit from using an HMM model that encodes the inherently continuous
dynamics of users' engagement states. Next, we included
cross-participant influence by using the CHMM and achieved a smaller
improvement in performance.  This is not surprising for a first
attempt; different pairs of conversational participants will have
somewhat different dynamics, and it is less likely that we could
completely encode this complex interaction using a simple,
speaker-independent model and limited training data. Considering that
the data we used are from spontaneous speech in real telephone calls
and the method does not encode any speaker-dependent information, the
results are reasonably good and promising.

\begin{table}[htb]
\centering
\caption{\label{tab:table2}{\em Engagement detection accuracy in continuous
  speech.}}
\begin{tabular}{|c|c|c|c|}
\hline 
  {\bf random} & {\bf isolated SVM} & {\bf HMM} & {\bf coupled HMM}  \\
\hline 
  {$(20\%)$} & {$47\%$} & {$61\%$} & {$63\%$} \\
\hline
\end{tabular}
\end{table}

\section{Conclusions}
In this work, we proposed to use affective information encoded in
speech to estimate users' engagement in computer-mediated voice
communications systems for mobile computing. To our knowledge, this is
the first work that attempts to estimate users' engagement in
spoken conversation. We tested this idea by developing a machine
learning system that can perform engagement detection in everyday
dialogue and achieved reasonably good results.

The main technical contribution of this work is a method for estimating users'
engagement based on a novel multilevel structure. Compared with
previous studies that focus on classifying user's emotional states
based on individual utterances, our method models the emotion
recognition and engagement detection problem in a continuous manner. In
addition, we encode the joint behavior of the participants in a
conversation using a CHMM. We demonstrated that our method
achieved much better results than the one based solely on low-level
acoustic signals of individual spoken utterances. A natural extension
of the study reported here is to extract additional types of affective information
from speech, such as linguistic information, and
then include them as the observation data of the high-level HMM to
improve the overall performance of engagement detection.

\bibliographystyle{IEEEtran}
\bibliography{IEEEabrv,emotion}
\end{document}